\documentclass[useAMS,usenatbib]{mn2e}
\usepackage{graphicx}
\usepackage{amsmath}
\usepackage{bm}
\usepackage{color}
\usepackage{url}
\usepackage{listings}
\usepackage{hyperref}

\newcommand{\modif}[1]{\textcolor{black}{#1}}

\title{Robust, open-source removal of systematics in Kepler data}

\author[S. Aigrain et al.]{S. Aigrain$^{1}$\thanks{E-mail:
    suzanne.aigrain@astro.ox.ac.uk}, H. Parviainen$^{1,2,3}$,
  S. Roberts$^{4}$, S. Reece$^{4}$ \& 
  T.\ Evans$^{1}$ \\
  $^{1}$Sub-department of Astrophysics, Department of Physics,
  University of Oxford, Oxford OX1 3RH, UK\\
  $^{2}$Instituto de Astrof\'isica de Canarias, V\'ia L\'actea s/n, 38205, La Laguna, Spain\\
  $^{3}$Departamento de Astrof\'isica, Universidad de La Laguna, 38205 La Laguna, Tenerife, Spain\\
  $^{4}$Machine Learning Research Group, Department of
  Engineering Science, University of Oxford, Oxford OX1 3PJ, UK}
\begin{document}

\date{Accepted \ldots Received \ldots; in original form \ldots}

\pagerange{\pageref{firstpage}--\pageref{lastpage}} \pubyear{2016}
\maketitle

\label{firstpage}

\begin{abstract}
 We present ARC2 (Astrophysically Robust Correction 2), an open-source \textsc{Python}-based systematics-correction pipeline to correct for the  \textit{Kepler} prime mission long cadence light curves. The ARC2 pipeline identifies and corrects any isolated discontinuities in the light curves, then removes trends common to many light curves. These trends are modelled using the publicly available co-trending basis vectors, within an (approximate) Bayesian framework with `shrinkage' priors to minimise the risk of over-fitting and the injection of any additional noise into the corrected light curves, while keeping any astrophysical signals intact. We show that the ARC2 pipeline's performance matches that of the standard \textit{Kepler} PDC-MAP data products using standard noise metrics, and demonstrate its ability to preserve astrophysical signals  
 using injection tests with simulated stellar rotation and planetary transit signals. Although it is not identical, the ARC2 pipeline can thus be used as an open source alternative to PDC-MAP, whenever the ability to model the impact of the systematics removal process on other kinds of signal is important.
\end{abstract}


\section{Introduction}

During \modif{a little more than 4 years} of operations, the Kepler space mission produced
continuous, high-precision light curves for over 150\,000 stars, with
a cadence of 29.4\,min. This forms a very rich dataset for a wide
range of exoplanet and stellar variability studies. However, the light curves also
contain instrumental artefacts and systematic trends. Correcting these while
preserving `real' astrophysical variability is challenging, but very
important for the community to make the most of the Kepler data. 

The publicly available Kepler data products
\citep{KeplerArchiveManual} include three versions of the time-series
data for each target star. The target pixel files contain flux
measurements in each of the individual pixels falling within a
pre-defined area around each star. These have been \modif{through low level CCD and instrument calibrations and pixel-level cosmic ray removal}, but are otherwise `raw'. The
light curve files contain two versions of the light curve, dubbed SAP
(Simple Aperture Photometry) and PDC (Pre-search Data
Conditioning). \citet{jen+10} and \citet{KeplerDataProcHandbook} give an overview
of the data processing steps involved in producing both the target
pixel files and the light curves. The SAP light curves are obtained by
summing the flux falling within a subset of the pixels included in the
target pixel files, and applying a correction for background flux
\citep{twi+10a}. The PDC light curves result from additional
processing steps designed to remove instrumental artifacts and
systematics \citep{stu+12a,smi+12}.  As its name indicates, the PDC
pipeline is primarily intended to ready the data for planetary transit
searches, and is not specifically optimized to preserve other forms of
astrophysical variability. The PDC data are nonetheless widely used
for both transit searches and variability studies \citep[for example stellar rotation studies,
see e.g.][]{rei+13,nie+13,mcq+13a,mcq+13b,mcq+14}, because they are
the only widely available set of light curves which a) cover the full target list and time coverage the \emph{Kepler} mission, and b) are free of many of the
systematics, which dominate the SAP light curves. However, some versions of the PDC pipeline were prone to over-correction (removal of real astrophysical
variability and injection of additional noise, see
e.g. \citealt{rob+13}). Additionally, the PDC pipeline is not in the public domain, making it difficult to reproduce results based on this pipeline independently, or to perform independent evaluations of the way in which it affects different types of stellar and planetary signals (as required, for example, for planet incidence studies). This motivated us to develop an alternative procedure to correct systematic trends and artifacts in \emph{Kepler} data, aiming to match (or improve on) the performance of the PDC pipeline, but with a specific
emphasis on retaining real astrophysical variability, and a commitment to making the code publicly available. 

The Kepler detector consists of 21 modules arranged in a $5 \times 5$
grid with missing corners, each containing two
$4{\rm K}\times 2 {\rm K}$ CCDs \citep{KeplerInstrumentHandbook}. The two
halves of each CCD are read out separately, leading to 4 output
channels per module. Throughout this paper, we refer to each module
plus output channel combination as a modout and identify it as $X.Y$
where $X$ is the module number (2 to 24) and $Y$ the output channel number (1 to
4). For the tests described in this paper, we focussed on 4 modules: 2.1, 7.3,
13.1 and 17.2. These were selected because they are collectively
representative of the range of systematic effects, which affect Kepler
data: 2.1 is located near the upper left corner of the field of view
and is particularly sensitive to focus changes, 7.3 is `atypically
typical', 13.1 is at the centre of the focal plane, and 17.2 contains
some peculiar image artefacts (J.\ Smith, priv.\ comm.). 

Every three months, the Kepler satellite rolls by $90^{\circ}$ about its boresight
in order to keep its solar panels pointing towards the Sun. Each
3-month period between rolls is known as a quarter, and we follow the usual convention of referring to each quarter as Q0, Q1, Q2, etc\ldots As each star
falls on a different modout in each quarter, the systematics observed
vary from quarter to quarter, and their treatment is best
carried out separately for each quarter. Each star returns to
approximately the same position on the detector every 4 quarters (1
year), so the systematics observed in a given star's light curve in
quarters (say) 3 and 7 tend to be mutually similar. 

\subsection{Systematics removal in the PDC pipeline}
\label{sec:pdc}

\begin{figure*}
  \centering
  \includegraphics[width=\linewidth]{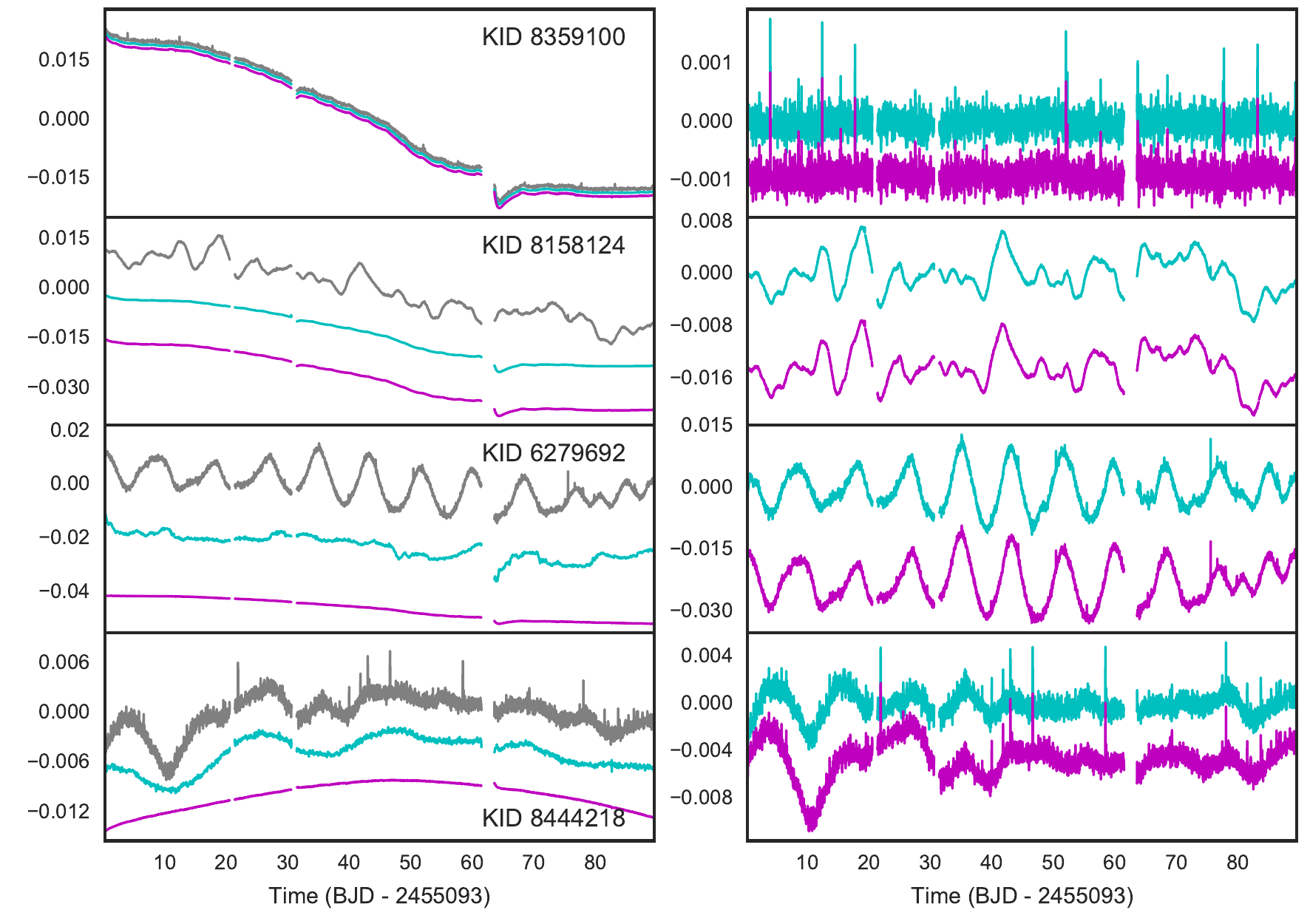} 
  \caption{Example light curves before and after systematics
    correction (random selection from Q3, modout 7.3). The left column shows the SAP light curve for each
    object in grey, with the corrections applied by the PDC-MAP and
    by our own pipeline in cyan and magenta, respectively. The right column
    shows the corrected light curves (PDC-MAP in cyan, this work in
    magenta). In both columns, arbitrary vertical offsets have been applied
    to separate the different curves. Both corrections are almost identical in the \modif{top} two cases, but the PDC-MAP correction adds considerably more broad-band noise and over-corrects the intrinsic variability in the bottom two examples (see text for details).}
  \label{fig:ex_pdc}
\end{figure*}

In this subsection, we give a brief overview of the systematics
removal methods implemented in the pipeline which produces the
publicly available PDC light curves. This is a very simplified
description, intended only to set the scene for the present paper, and
a number of important features have been omitted for the sake of
brevity; full details are given in the relevant publications.

The standard approach for systematic trend removal in transit survey
data is to model each light curve as a linear combination of
systematic trends, which are derived either from ancillary engineering
and meteorological data (such as telescope pointing, focus, seeing and
airmass, see e.g. \citealt{bak+07}) or from the light curves
themselves \citep{tam+05,kov+05}. In the latter case, variants of
Principal Component Analysis (PCA) are used to construct a reduced
basis from the light curves.

\begin{figure*}
  \centering
  \includegraphics[width=\linewidth]{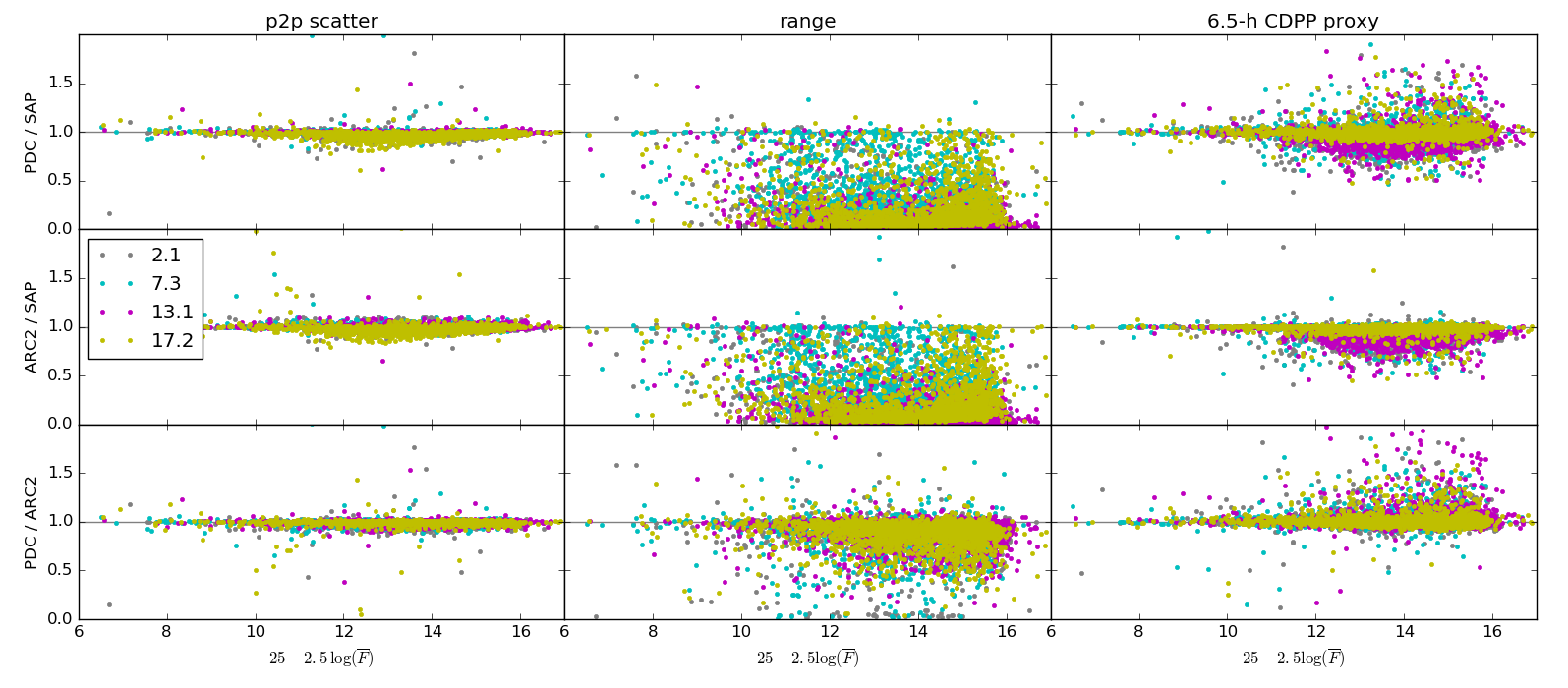} 
  \caption{Statistical comparision of the SAP, PDC-MAP and ARC2 light
    curves for Q3. The left, middle and right columns show the short-term scatter, range and 6.5-h CDPP proxy, respectively, while the top, middle and bottom rows show the ratio of these quantities between PDC and SAP, ARC2 and SAP, and PDC and ARC2 light curves, respectively. 
    In all cases, the $x$-axis shows a magnitude-like
    quantity based on the median flux in each light curve, but with a very approximate zero-point. The
    different colours correspond to different modouts. The detailed behaviour of the different modouts is discussed
  in Section~\ref{sec:comp}.}
  \label{fig:sig_ran_comp}
\end{figure*}

The PDC pipeline broadly follows this paradigm. In its original
version \citep{twi+10a,twi+10b}, known as PDC-LS (for least squares),
the basis was constructed from ancillary engineering data and the
coefficient relating each systematic trend (each term in the basis) to
each light curve were derived by least-squares (or maximum likelihood)
fitting. However, this method suffered from two problems: overfitting
(removal of real astrophysical variability) and injection of noise
into the light curves (a side-effect of overfitting combined with
noisy basis vectors). 

To address these issues, a new version of the PDC was introduced,
known as PDC-MAP (for maximum a posteriori, \citealt{smi+12}). In the
PDC-MAP, the systematic trends (which are known, in Kepler jargon, as
co-trending basis vectors, or CBVs), are computed from the light curves
themselves, by applying singular value decomposition (SVD) to the 50\%
of the light curves which show the strongest mutual correlation. This
is done separately for each of the 4 output channels in each of the 21
modules composing the Kepler detector. The
coefficients linking each CBV to each light curve are then evaluated
in a two-step process. First, preliminary estimates are computed using
the same least-squares method as in PDC-LS. The resulting coefficients
are used to construct prior distributions which are parametrised
according to stellar magnitude and position on the sky, and the final
coefficients are found by maximising the marginal likelihood for each
star, subject to these priors.

The PDC-MAP pipeline performs significantly better than the PDC-LS one, and it is
less susceptible to over-fitting and noise injection, although these
effects do remain apparent in some of the light
curves. To address the small number of cases where the PDC-MAP pipeline performance was not entirely satisfactory, a third version, known as PDC-msMAP (``ms" stands for ``multi-scale"), was developed by \citet{stu+14}. This version uses a wavelet transform to split the light curves into three channels, or bands, which are processed separately, resulting in a more effective separation of intrinsic and common-mode
signals, and improved removal of the systematics. The final version of the \emph{Kepler} data \modif{(\href{https://archive.stsci.edu/kepler/release_notes/release_notes25/KSCI-19065-002DRN25.pdf}{Data Release 25})}, now available at MAST, uses the PDC-msMAP pipeline, and will be used for comparison throughout this paper, though for the sake of brevity, we hereafter refer to the PDC-msMAP pipeline and data using the shorthand `PDC'.

Figure~\ref{fig:ex_pdc} shows a few representative
examples of SAP light curves from Q3 (in grey), together with the correction applied by the PDC pipeline, and the resulting PDC light curve (in cyan). (The magenta curves refer to the new ARC2 correction, which is presented later in this paper.) The top two rows show a relatively quiet star and a short time-scale ($<5$\,d) variable, respectively. In those cases the PDC pipeline performs as well as one could expect. The bottom two rows show two longer time-scale ($>5$\,d) variables. Those illustrate the limitations of even the latest version of the PDC pipeline. \modif{The correction adds broad-band noise into the light curves, and over-corrects the intrinsic variability. Of course, we do not know the ground truth, since these are real examples, but we can say this with some confidence, because the correction is unlike that applied to the vast majority of other light curves, but instead mimics variations in the individual light curves, which are typical of evolving, rotating star spots.} One of the design considerations for our own pipeline was to try to reduce the incidence of these problems even further, without sacrificing the otherwise excellent overall performance.

The overall performance of both the PDC and our own pipeline are illustrated in a statistical manner in the top panel of Figure~\ref{fig:sig_ran_comp}. This figure shows various estimates of the light curve scatter on different timescales as a function \modif{of} magnitude, in the form of ratios between the PDC, SAP and our own correction, for 4 representative channels in Q3. The left-hand column shows the point-to-point (p2p) scatter $\sigma$, measured as the standard deviation of
the first difference of each light curve. Using the first difference ensures that any
long-term trends are taken out of the equation, giving an estimate of the high-frequency noise level alone. The middle column shows the range $R$, first introduced by \citet{bas+10a,bas+10b} as a relatively noise-independent measure of light curve amplitude, and defined as the interval
between the $5^{\rm th}$ and the $95^{\rm th}$ percentiles of the
normalised\footnote{Throughout this paper, all
light curves are assumed to be normalised by dividing them by the
median flux value.} flux values. $R$ captures variations on a wide range of
time-scales, including both systematics and intrinsic
variability. Finally, the right-hand column shows the 6.5-h combined differential photometric precision (CDPP)\footnote{Throughout this paper we use our own proxy estimate of the CDPP, as defined in \citet{aig+16}. The true CDPP is estimated by the \emph{Kepler} transit search pipeline and published together with the PDC light curves, but we have chosen to compute our own proxy for it so that it can be compared directly to the same quantity for the light curves processed with our own pipeline.} which is widely used to evaluate the noise level on transit timescales. 

At the time this work was initiated, the PDC-msMAP results were not yet available for most quarters, and it was noticeable that the PDC-MAP pipeline introduced significant amounts of high-frequency noise in a relatively large fraction of the light curves. The top-left panel of Fig.~\ref{fig:sig_ran_comp} shows that this problem has been significantly reduced in the PDC-msMAP pipeline. On average, the latter leaves
$\sigma$ unchanged, and even reduces it in certain
channels, with only a few individual exceptions. As most light curves are initially systematics-dominated, so
the PDC correction should -- and does -- reduce $R$ significantly
in most cases, as illustrated in the top-middle panel, but it is hard to tell whether any of the reduction in $R$ results from the removal of real variability. Finally the top-right panel shows that the 6.5~h CDPP is more or less unaffected on average, and even reduced slightly in some channels, but it is interesting to note that it is significantly increased (by 10\% or more) for a non-negligible fraction of the cases (about 10\% for these 4 particular channels in Q3). 

\subsection{An open-source systematics removal pipeline matching the PDC performance}

As explained above, we seek to develop an alternative systematics-correction pipeline for \emph{Kepler}, which comes as close as possible to matching the excellent performance of the PDC pipeline, but for which the code and the input data are all publicly available, to enable independent estimates of the impact of this pipeline on different kinds of astrophysical signals of interest. Furthermore, it would be desirable to reduce the extent to which longer-term astrophysical variability is affected, and further minimise the injection of short-term noise into the corrected light curves.

In \citet[][hereafter Paper I]{rob+13}, we presented an alternative method to
identify and correct common-mode systematics (trends present, to a
greater or lesser extent, in the majority of light curves), which we
called the ARC (Astrophysically Robust Correction) method, as it was
specifically designed to minimize the risk of removing or altering
astrophysical variability along with the instrumental
systematics. Like the PDC, the ARC is based on decomposing each light
curve into a linear superposition of systematic trends, plus a unique
vector representing intrinsic variability and random noise. Like
in the PDC-MAP and msMAP versions, the systematic trends are identified from the light curves
themselves, for each output channel, although the procedure for doing so is slightly different (the ARC uses an information entropy criterion to identify genuinely systematic trends). Once identified, the trends are
smoothed before they are applied to the individual light curve. Finally, adaptive,
zero-mean (shrinkage) priors were used to evaluate the coefficients
linking each systematic trend to each light curve, in an effort to reduce the risk of over-fitting. 

Since Paper I, which included tests on Q1 data only, we have applied the ARC to later quarters, and it has become apparent that \modif{the} ARC trend identification method results in basis vectors that are usually extremely similar to the CBVs. We thus concluded that it may be more expedient to use the published CBVs than to derive our own (specially as the trend identification is by far the most CPU intensive part of the ARC pipeline). Using the same basis vectors as the PDC also maximises the chances that the light curves corrected with our pipeline are as similar as possible to the PDC version, whenever that is desirable. 
Furthermore, the smoothing method used in Paper I (empirical mode decomposition, \citealt{hua98}), cannot adequately reproduce the discontinuities and sharp decays that affect essentially all light curves, following each (approximately monthly) data down-link event. On the other hand, \modif{during our initial tests of the ARC on quarters 2 to 5,} we noted that both the first few CBVs, and the raw ARC basis vectors, contain some high-frequency structure, which appears to be
real, in the sense that it is also present in the light curves of
bright stars (where the photon noise does not mask these effects). \modif{The origin of this high-frequency structure is not clear, but its existence} implies that smoothing the basis vectors may not be a good idea after all.

This leaves us with the trend removal section of the ARC, which is essentially an alternative to the MAP portion of the PDC-MAP process. Both pipelines use priors over the coefficients relating each basis vector to each light curve, but instead of constructing those by parametrising the distribution of these coefficients as a function of (e.g.) a star's magnitude and location on the detector, the ARC approach consists in starting with a zero-mean Gaussian prior over each coefficient, and applying a Gamma hyper-prior on the width of that Gaussian. Unless there is particularly strong support in the data for a particular basis vector, the prior for the corresponding coefficient will rapidly shrink to zero (hence the name ``shrinkage prior"). This is intended to minimise the risk of over-fitting, which leads to the noise-injection and variability suppression problems we noted in some PDC light curves. This trend removal procedure is implemented in a
variational inference framework (see the Appendix of paper I for full details), which is
extremely fast: given a set of basis vectors,
correcting a given light curve for a given quarter takes of order a second, and all the light curves (for all channels) less than half an hour on our 96-core linux cluster. 

\modif{Our new pipeline is thus essentially a stripped down version of the original ARC, retaining the variational Bayes framework and shrinkage priors, but using the pre-computed CBV basis vectors. Prior to the CBV correction, however, we must include a jump-correction step. Many light curves contain one or two isolated jumps, or 
discontinuities, which are suspected to be due to individual pixel malfunction. They are identified and corrected as part of
the PDC pipeline, but the published data products do not provide enough
information to separate this from the systematic trend correction, so
it was necessary for us to develop our own jump correction. Together, our new jump correction and CBV correction method form the ARC2 pipeline.}

The rest of this paper is structured as follows.  \modif{In Section~\ref{sec:jumps}, we introduce and test our new jump correction correction. In
Section~\ref{sec:cbv} we describe our method for applying the
CBVs to the light curves, and present} a simple way of
deciding, for each light curve, how many CBVs should be used in the
correction. This Section also includes links to the corrected data and
the code used to produce them. Finally, in Section~\ref{sec:tests} we evaluate the performance of
our method using injection tests, and compare it to the PDC
method.

\section{Detection and correction of individual discontinuities}
\label{sec:jumps}

Prior to correction using CBVs, it is vital to remove and correct for the numerous discontinuities present in the light curves.
This sudden pixel sensitivity dropout (SPSD) correction is divided into: a) detection, b) classification, and c) correction of discontinuities. The detection phase identifies discontinuities in the light curve, the classification phase identifies the type of the discontinuity (SPSD, flare, or transit), and the correction phase corrects the SPSDs but leaves flares and transits untouched. Each of these steps is detailed below.

\textcolor{black}{Simple approaches to discontinuity detection, such as convolving a step function with the light curve (matched filter) or looking for outliers in the difference between consecutive data points (or between every other data point) perform very well in white noise and provided there are very few data gaps. However, we were keen to develop a method that performed well even for light curves containing significant amounts of short-term variability. We therefore opted for an approach based on Gaussian Process regression with a change-point built into the covariance matrix. Early in the process of developing this method, we carried out simple tests on example light curves, which confirmed that this approach outperforms the matched filter and first- or second difference approaches for variable stars, while performing equivalently well for quiet stars.}

The discontinuity detection is \modif{based on a moving window likelihood ratio test between with- and without-discontinuity GP models.} 
The light curve, $\mathbf{f}$, is modelled as a zero-mean Gaussian Process (GP) with a covariance matrix $\mathbf{K}$, that is
\begin{equation}
    \mathbf{f} \sim \mathcal{N}(0, \mathbf{K}).
\end{equation}
The covariance matrix elements are defined by a covariance function (kernel) with the light curve cadences $\mathbf{c}$ as the input parameter.
\modif{We use two kernels, without and with a breakpoint ($\mathbf{K_0}$ and $\mathbf{K_1}$, respectively), defined as
\begin{equation}
    \mathbf{K_{0,ij}} = a^2 \exp\left(-\frac{|c_i - c_j|}{\lambda}\right) + \sigma^2 \delta_{ij},
\end{equation}
\begin{equation}
    \mathbf{K_{1,ij}} = a^2 \exp\left(-\frac{|c_i - c_j|}{\lambda}\right) \times B(\beta, c_i, c_j) + \sigma^2 \delta_{ij}.
\end{equation}
The kernels represent the light curve as a sum of an exponential kernel with an output scale $a$ and input scale $\lambda$
and an average white noise term. The with-breakpoint kernel, $\mathbf{K_1}$, includes a  breakpoint function 
\begin{equation*}
    B(\beta, c_i, c_j) = \begin{cases}
    1 &\text{if } \left( c_i \leq \beta \land c_j \leq \beta\right) \lor (c_i \geq \beta \land c_j \geq \beta)\\
    0 &\text{else}
    \end{cases}
\end{equation*}
that forces the covariances between the points on the different sides of a given breakpoint cadence, $\beta$, to zero.}

\begin{figure}
 \centering
 \includegraphics[width=\columnwidth]{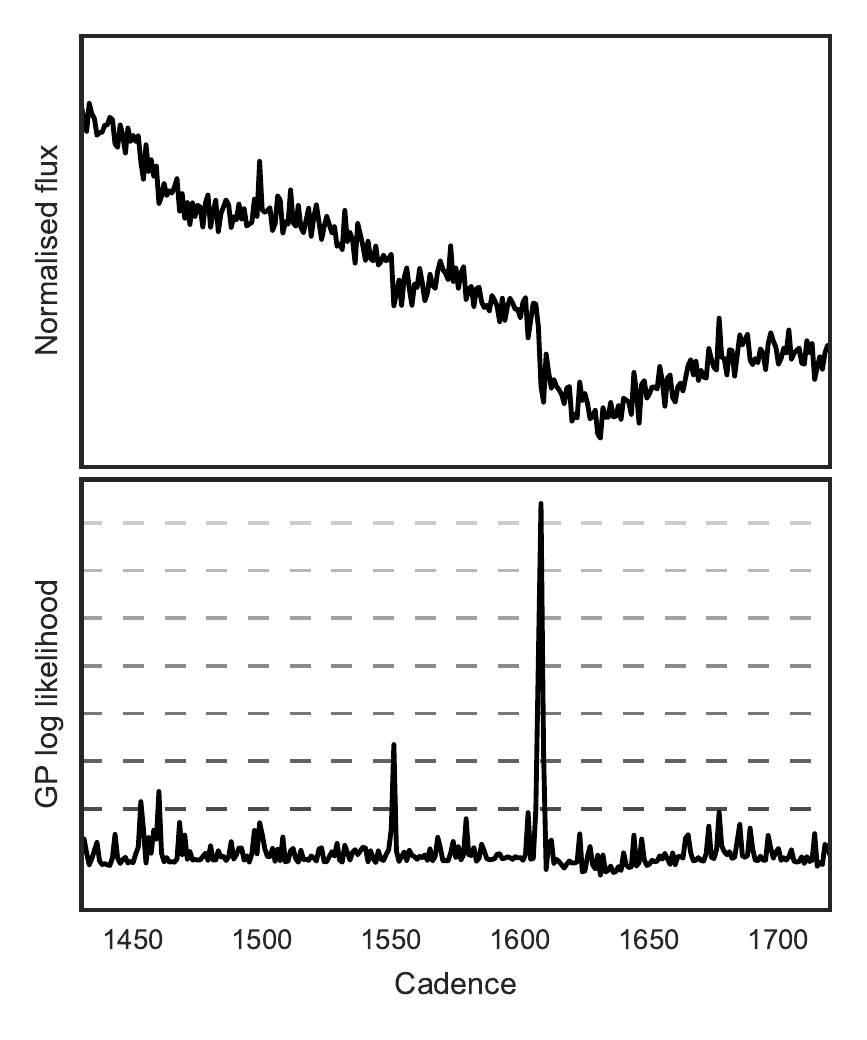}
 \caption{An example illustrating the discontinuity search: 
  a light curve containing long-period variability, white noise,
  correlated noise, outliers, and two SPSD signals (top),
  \modif{and the GP log likelihood difference between the with-breakpoint and without-breakpoint GP kernels,}
  and the 10--80 MAD (median absolute deviation) levels for the log likelihood distribution 
  marked as horizontal slashed lines (bottom).}
 \label{fig:jump_identification}
\end{figure}

\modif{The discontinuity search starts with the removal of strong individual outliers identified using a narrow
running median filter. Next, we learn the $\mathbf{K_0}$ hyperparameters $a$, $\lambda$, and $\sigma$
by maximising the GP likelihood for a subset of the light curve. Then, we fix the hyperparameters to the optimised values, 
and calculate $\mathbf{K_1}$ log likelihoods, $\log \mathcal{L}_1$, for each $\beta = [0..N]$ using a $n_\mathrm{p}$ cadence-wide moving window 
centred around $\beta$ (where $n_\mathrm{p}=150$ by default). We also calculate $\mathbf{K_0}$ log likelihoods, $\log \mathcal{L}_0$, for each
window, and subtract these from $\log \mathcal{L}_1$ to obtain a series of log likelihood ratios, $\log (\mathcal{L}_1/\mathcal{L}_0)$. 
Finally, we identify discontinuities as positive outliers in the log likelihood ratio,} as illustrated  in Fig.~\ref{fig:jump_identification}.

\begin{figure*}
 \centering
 \includegraphics[width=\textwidth]{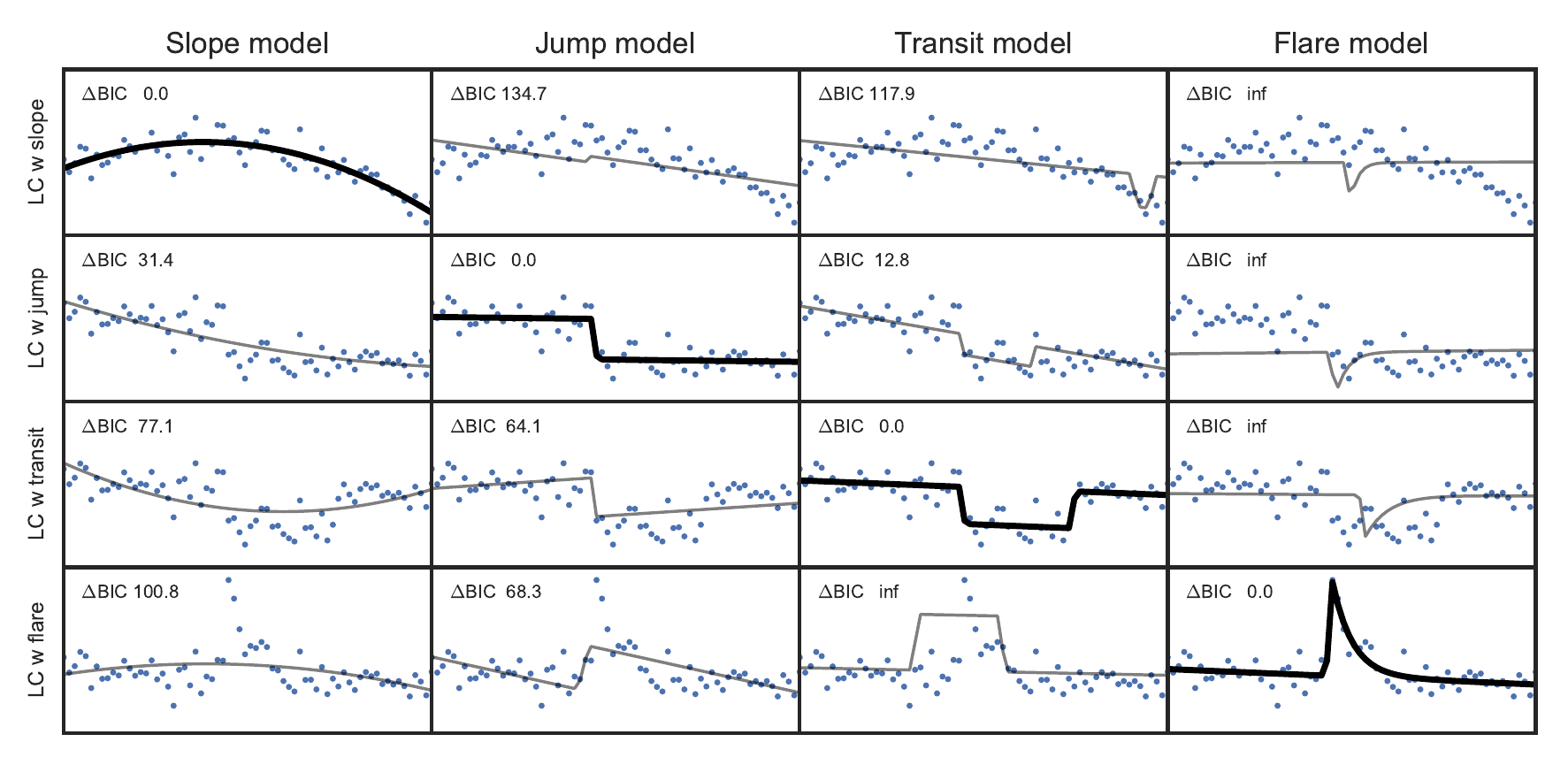}
 \caption{Discontinuity classification. Each row shows a light curve with a discontinuity (blue dots), and each column shows a discontinuity model fitted to the light curve (black line) with the model's BIC value relative to the lowest BIC value of the fitted models. The model with the lowest BIC value (thick black line) is chosen to represent the discontinuity type. The infinite $\Delta$ BIC values arise when the priors for the model parameters reject the model directly, such as with flares with negative amplitude.}
 \label{fig:jump_classification}
\end{figure*}

The Kepler light curves contain several types of signals that can cause rapid
changes in the flux. Besides the instrumental SPSDs that we want to find and correct, 
we also have astrophysical signals---such as planetary transits, binary eclipses, and stellar 
flares---that we do not want to remove. Thus, an automated SPSD correction routine
needs a way to distinguish an SPSD from these other discontinuities. The discontinuity classification uses a 
Bayesian information criterion (BIC)-based model selection approach to select between a set of possible discontinuity models. The current code implements five models: a) false alarm modelled by a low-order polynomial, b) transit-like, c) flare, d) SPSD, and e) SPSD followed by an exponential drift to a new level. The discontinuity classification phase fits the models to each discontinuity, modelling the flux baseline either with a low-order polynomial or a
Gaussian Process, and selects the model with the lowest BIC value as the true model, as shown in Fig.~\ref{fig:jump_classification}.

Finally, in phase c), we correct the identified SPSDs based on the fitted discontinuity model and
save the information about the identified discontinuities for later use. 

\section{Optimized use of the CBVs}
\label{sec:cbv}

\subsection{CBV fitting using Variational Bayes}

We fit each light curve using the standard linear basis model:
\begin{equation}
F^{(i)}_j = \sum_{k=1}^K w^{(i)}_k C_{kj} + \epsilon^{(i)}_j
\end{equation}
where $F^{(i)}_j$ is the flux measured for star $i$ in observation
$j$, $C_{kj}$ is the value of the $k^{\rm th}$ basis vector
(systematic trend, or CBV) in observation $j$, $w^{(i)}_k$ is the
coefficient, or weight, relating basis vector $k$ to light curve $i$,
and $\epsilon^{(i)}_j$ represents the residuals of the correction for
star $i$ in observation $j$. In the remainder of this section, we omit
the superscript $^{(i)}$ for simplicity -- the analysis is done
separately for each light curve. Note that $\epsilon$ contains both
intrinsic variability and noise, representing the total residual for the
purposes of the systematics correction.

In a simple least-squares framework (such as PDC-LS), one seeks the
set of $w$'s which minimises the total squared residuals, $\sum_j \epsilon_j^2 $. If the residuals are assumed to be drawn
independently from a Gaussian distribution with known precision (inverse
variance) $\beta$ this is equivalent to
maximising the likelihood of the model:
\begin{equation}
p(\mathbf{F}|\mathbf{w},\mathbf{C}) =
\mathcal{N}(\boldsymbol{\epsilon};0,\beta^{-1} \mathbf{I}).
\end{equation}
(Note that strict equivalence only holds if $\beta$ is known, if it is
a free parameter then the full likelihood expression must be used.)

Maximum likelihood linear basis models are notoriously prone
to over-fitting. Better results can be obtained in a Bayesian
framework, by using
priors over the $w$'s to encapsulate any external information
available over the expected values for the weights, and maximise the
posterior distribution instead of the likelihood. 
\begin{equation}
p(\mathbf{w}|\mathbf{F},\mathbf{C}) =
\frac{p(\mathbf{F}|\mathbf{w},\mathbf{C})  \, p(\mathbf{w})}{\int
  p(\mathbf{F}|\mathbf{w},\mathbf{C}) \, p(\mathbf{w}) \, {\rm d}\mathbf{w}},
\end{equation} 
where the normalisation constant in the denominator is the model
evidence $p(\mathbf{F}|\mathbf{C})$. In the PDC-MAP pipeline, the priors over
the $w$'s are based on the distribution of the coefficients derived in
the maximum likelihood case (i.e.\ in the absence of priors),
parametrised as a function of star position and magnitude. This
reflects the belief that stars which are near each other on the
detector and have similar brightnesses should also display similar
systematics. As we have seen, it does reduce the overfitting problems
which had been noted in PDC-LS, but does not entirely do away with
them. One plausible explanation for this is that the PDC-MAP priors
themselves are affected by overfitting in the initial, maximum
likelihood step. Furthermore, a fixed prior, as in the MAP model,
does not guarantee model shrinkage as required to avoid over-fitting. 

To reduce the risk of over-fitting further we perform inference using Bayesian learning, which allows us to regularise the model using priors
which specifically penalise larger weights, and make it less likely
that one basis vector will compensate for another. A natural choice
for this is to use zero-mean Gaussian priors:
\begin{equation}
p(w_j|\alpha_j) = \frac{\alpha_j}{\sqrt{2 \pi}} \exp\left(-\alpha_j w_j^2 / 2 \right)
\end{equation}
for each $j$ (the individual prior weights are treated as
mutually independent).  Furthermore, we do not fix the priors, but
instead treat the inverse variances, $\boldsymbol{\alpha} = \{\alpha_j\}$, as
parameters themselves, subject to their own prior
$p(\boldsymbol{\alpha})$, for which we use a Gamma
distribution\footnote{Our choice of priors over $\mathbf{w}$ and
  $\boldsymbol{\alpha}$ is also mathematically convenient, because
  they are conjugate with each other and with the likelihood (which is
  Gaussian), enabling a number of the integrals involved in the
  inference to be performed analytically.}. Unless there is strong evidence for a non-zero weight
for a particular light curve / basis vector combination, the Gamma prior over
$\alpha$ will tend to make the distribution over $w$ collapse close to a delta
function centred on zero, so most basis vectors will have zero weight in most
light curves. This is often referred to as
automatic relevance determination (ARD) or shrinkage.

We then seek to evaluate the posterior distribution over the weights
$\mathbf{w}$, marginalised over the prior precisions
$\boldsymbol{\alpha}$:
\begin{equation}
p(\mathbf{w}|\mathbf{F},\mathbf{C}) \propto \int
p(\mathbf{F}|\mathbf{w},\mathbf{C})  \, p(\mathbf{w}|\boldsymbol{\alpha}) \, p(\boldsymbol{\alpha}) \, {\rm d}\boldsymbol{\alpha},
\end{equation} 
and to maximise it with respect to $\mathbf{w}$. In general, the
posterior distribution is unknown and is not analytically
tractable. Numerically evaluating and optimizing the posterior would
require evaluating the likelihood over a very large number
of $(\mathbf{w},\boldsymbol{\alpha})$ combinations, which is
unfeasible, especially as it needs to be done for every Kepler
light curve. 

An elegant workaround consists in approximating the posterior with a
proposal distribution which is analytically tractable, and iteratively
refining the latter so that it approaches the true posterior. This
class of methods is known as approximate inference. More specifically,
one can restrict oneself to proposal distributions which belong to the
exponential family: refining the proposal then consists in optimising
an integral with respect to a functional, which is typically done
using the calculus of variations. This approach is thus known as
variational inference, or variational Bayes (VB). A detailed
description of the VB method as applied to our linear basis model was
given in the appendix of Paper I, so we do not repeat it here. The
algorithm essentially consists in cycling through a set of update
equations for the weights $\mathbf{w}$, the prior precisions
$\boldsymbol{\alpha}$ and the noise precision $\beta$. Importantly,
the model is guaranteed to improve each iteration, thus providing robust
convergence which typically occurs after just a few 
iterations. The computational requirement scales as $JK^2$, where $J$
is the number of observations and $K$ is the number of basis
vectors. Our {\sc Python} implementation of the method runs in a fraction of a second per light curve per quarter ($\sim 4300$ observations) for up to 8 CBVs. 

\subsection{How many CBVs?}

\begin{figure*}
  \centering
  \includegraphics[width=0.49\linewidth]{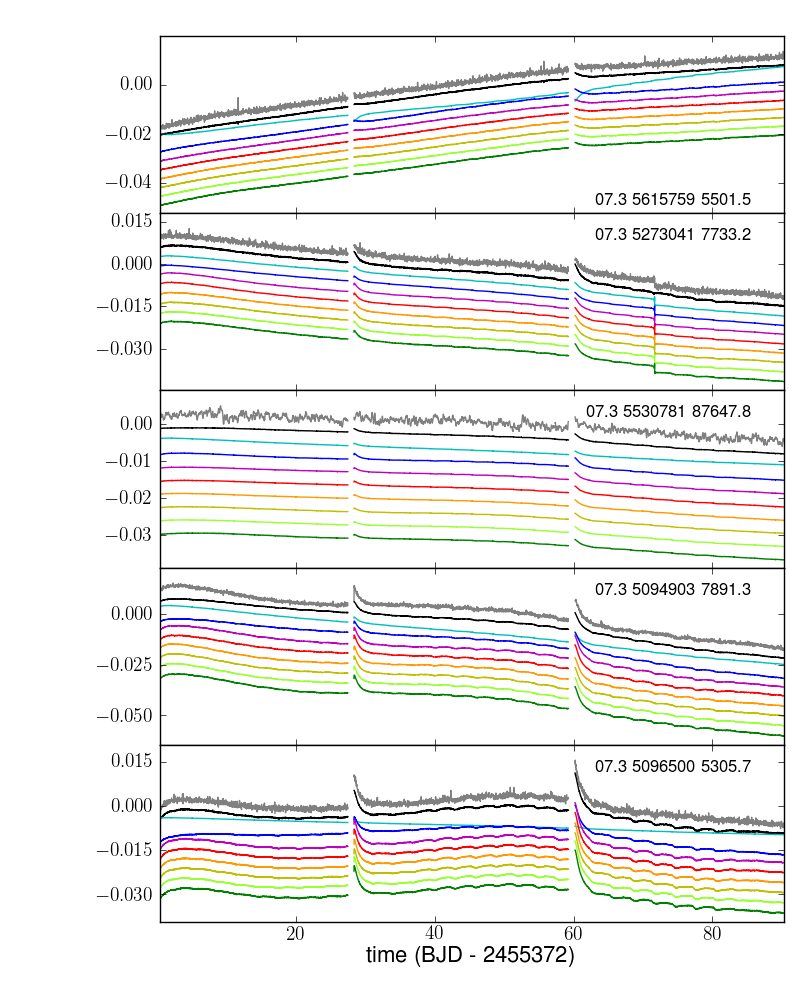}
  \includegraphics[width=0.49\linewidth]{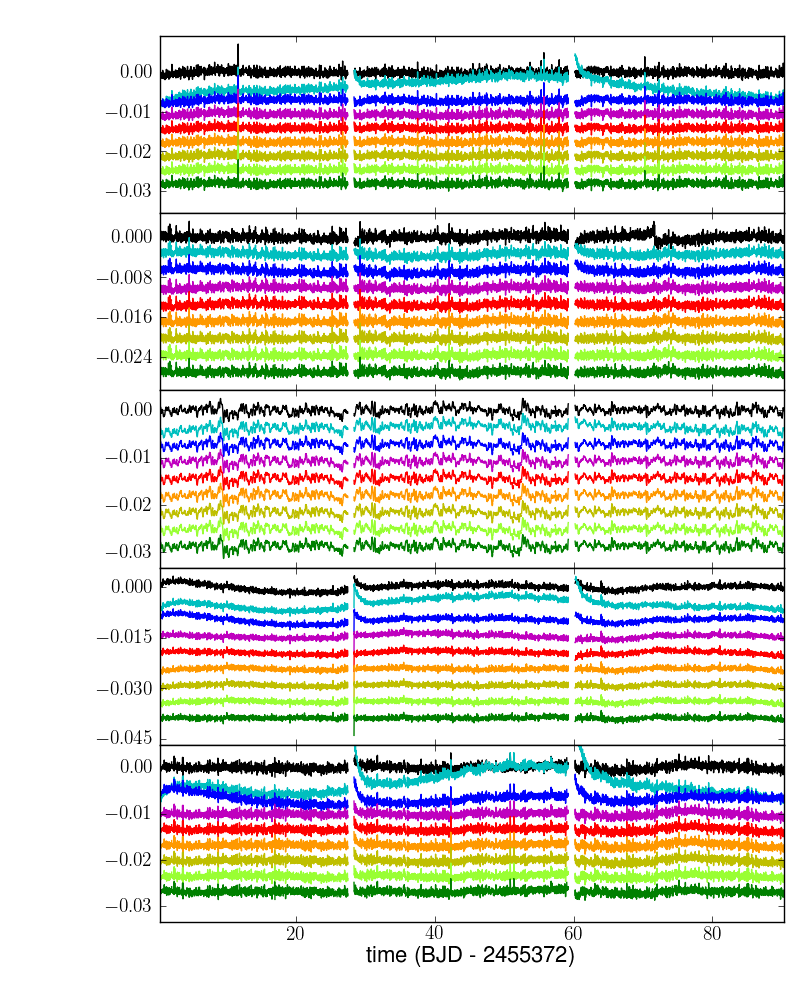}
  \caption{Comparison of the systematics correction using 1 to 8 CBVs,
    for a random selection of stars in Q3, modout 17.3. The left panel
    shows the original (SAP) light curve in grey and the corrections
    applied using 1 to 8 CBVs in different colours (cyan to green, and
    top to bottom). The PDC-MAP correction is also shown in black, for
    comparison. The right panel shows the corrected light curves,
    using the same colour-coding. In both panels, the different light
    curve versions have been offset by an arbitrary amount, for
    clarity.}
  \label{fig:ncbv_ex}
\end{figure*}

Despite the measures described in the previous section to minimise the
risk of over-fitting, the results still depend, in some cases, on the number of basis vectors
used. \modif{If our idealised model was correct, i.e.\ if each light curve was just a linear combination of the CBVs used, plus white Gaussian noise, and contained no intrinsic variability), this should not happen: the weights of any irrelevant CBVs included in the calculation should shrink to zero automatically. This doesn't happen in practice because the  intrinsic variability of many of the stars, which is not included in our systematics model (or in that of the PDC-MAP, of course), is significant at the \emph{Kepler} precision. To our knowlesge, there is no simple way of overcoming this problem, and one is forced to resort to more ad-hoc criteria.} 

The CBVs result from a singular value decomposition of a
subset of the light curves, and therefore the first CBV represents a
larger fraction of the overall variance of the light curves, and so
on. The PDC pipeline saves 16 CBVs but only uses at most 8 to
perform the correction. Additionally, a signal-to-noise ratio (SNR)
criterion is used to exclude CBVs which are deemed to contribute more
noise than useful information (see \citealt{smi+12} for details), so
the actual number of CBVs used ranges from 5 to 8. 
As the variational Bayes method is very fast, it is feasible to run it for every plausible
number of CBVs, $K$ (from 1 to 8). We first did this for a few
representative quarters (3 to 6) and output
channels, and performed a visual comparison of the results on a random
selection of light curves. 

We plotted the original (SAP) light curve, the correction
applied and the corrected light curve, for a few tens of stars
selected at random in each modout. A few examples are shown in
Figure~\ref{fig:ncbv_ex}. In most cases, the overall
shape of the correction is fairly insensitive to the number of CBVs
used, so long as it is at least 2 or 3. On the other hand, as the
CBVs are increasingly noisy, using more CBVs introduces more noise
into the light curves particularly for the brighter stars. It is
therefore important to use the smallest number of CBVs which gives an
adequate correction. Importantly, the examples we examined 
suggest that this number differs from star to star: in some cases,
including the $4^{\rm th}$ and $5^{\rm th}$ CBVs removes features
which appear systematic in nature, and which were not removed when
using only the first two. In other cases, even using the $2^{\rm nd}$ CBV
appears to have a detrimental effect. \footnote{\textcolor{black}{We also experimented briefly ith with using the CBVs in a different order than the one in which they are supplied, which results from the SVD process. However, this did not lead to any clear performance improvement. As our method is already designed to suppress irrelevant CBVs if the basis set supplied contains any, so the order of the CBVs should not matter, provided that all the relevant ones are included. }}

\begin{figure}
  \centering
  \includegraphics[width=\linewidth]{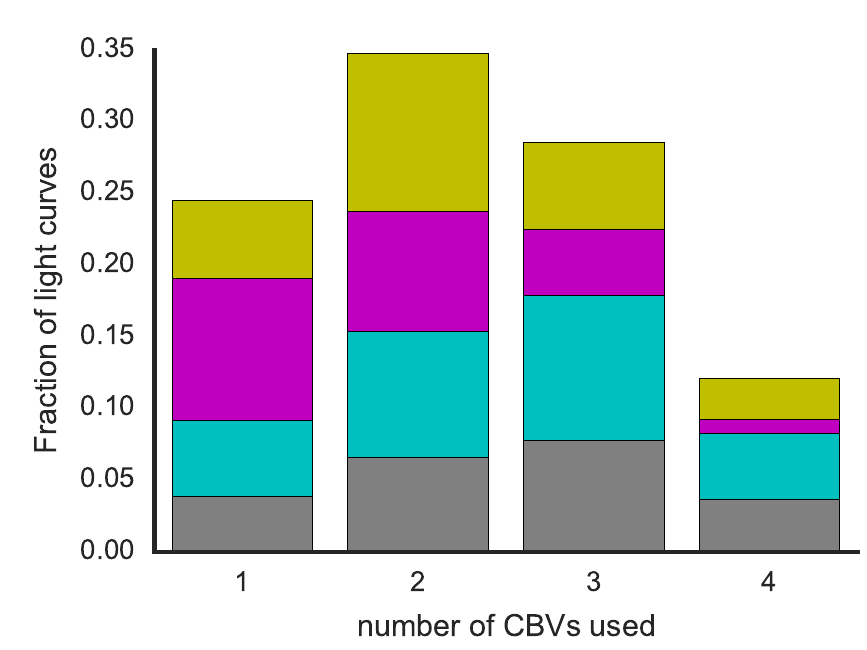}
  \caption{Distribution of the number of CBVs used for the 4 channels from Q3 shown in Figure~\ref{fig:sig_ran_comp} (same colour-coding).}
  \label{fig:ncbv}
\end{figure}

Rather than specifying an overall value for $K$ for, say, each
quarter and modout, we therefore decided to select the optimal number
of CBVs to use \textit{a posteriori}, on a light curve by light curve basis,
based on a statistical comparison of the light curve properties before
and after correction, using the simple statistics $R$ and $\sigma$. If most light curves are
initially dominated by systematics, the dependence of $R$ on $K$
tracks the extent to which the correction of systematics is improved
by adding more CBVs. Typically, $R$ decreases rapidly for low $K$ but
then reaches a plateau for larger $K$. Within this plateau regime, increasing
$K$ further does not usually improve the correction
significantly but it does introduce more noise, so one might simply
opt for the lowest value of $R$ at which the plateau is
reached. We initially define $K_{\rm opt}$ as the smallest value of $K$ for which $R(K)
< \langle R \rangle + 3 \sigma_R$, where $\langle R \rangle$ and
$\sigma_R$ are the median and standard deviation of $R$ over all
values of $K$, for a given light curve. 

However, when the initial amplitude of systematics is small relative
to that of the intrinsic variability, this criterion alone is
insufficient: $R$ can decrease with increasing $K$ not because more
systematics are being removed, but because real variations are
actually being removed. This is particularly frequent for bright,
variable stars. One way of testing for this would be to inject
realistic simulated variability signals into the light curve before
correction, and check how well they are recovered
post-correction. However, doing this for every light curve would be
prohibitively expensive. We do use such injection tests to
evaluate the overall ability of our correction method to preserve
astrophysical signals, but only on a subset of the data (see
Section~\ref{sec:inject}).

Fortunately, there is an easier way to
identify these problematic cases: visual inspection shows that, when
the correction removes what looks like real variability, it also
introduces significant amounts of high frequency noise into the light curves. This can be
diagnosed by tracking the dependence of $\sigma$ on $K$. Specifically, if
$\sigma(K_{\rm opt})/\sigma_0>1.1 $, where $K_{\rm opt}$ is determined
from the behaviour of $R$ as described above, and $\sigma_0$ is the
short-term scatter before correction, $K_{\rm opt}$ is decreased further
until the scatter ratio falls below the 1.1 threshold. The choice of
threshold is somewhat arbitrary, but its exact value is not
critical: when it is exceeded it is usually by a fairly wide margin.

Figure~\ref{fig:ncbv} shows histograms of the number of CBVs used for the 4 representative output channels from Q3 which featured in Figure~\ref{fig:sig_ran_comp}. Note that the number of CBVs used is never less than 1 or more than 4, and is fairly evenly distributed between those two values, but the distribution does vary noticeably from one channel to the next -- as do the CBVs themselves.

\subsection{Publicly available code}

The code is distributed as a GPL-licenced Python package \textsc{OxKeplerSC}\footnote{Available from \url{https://github.com/OxES/OxKeplerSC}.}. \modif{The package} can be installed using standard \textsc{Python} package installation procedures, and will be made available through \textsc{pip} in the future. The code is released as open source to encourage contributions and adaptations to specific use-cases.

After installation, the jump detection and detrending steps can be carried out for a single file or a whole directory as
\begin{lstlisting}
  keplerjc path_to_file_or_dir
  keplersc path_to_file_or_dir cbvdir
\end{lstlisting}
where, if given a directory, the user can optionally select a subset of modules, outputs, or their combinations to process using command line arguments. The discontinuity detection and correction takes several tens of seconds per light curve, and the detrending seconds. Batch processing of files is parallelised using MPI automatically if the scripts are called with \textsc{mpirun} or \textsc{mpiexec}.\footnote{The parallelisation depends on the \textsc{MPI4Py} package, and will be changed to use a higher-level implementation with \textsc{IPython} Parallel in the future.}

\section{Performance tests}
\label{sec:tests}

\subsection{Injection tests}
\label{sec:inject}

\begin{figure*}
  \centering
  \includegraphics[width=0.49\textwidth]{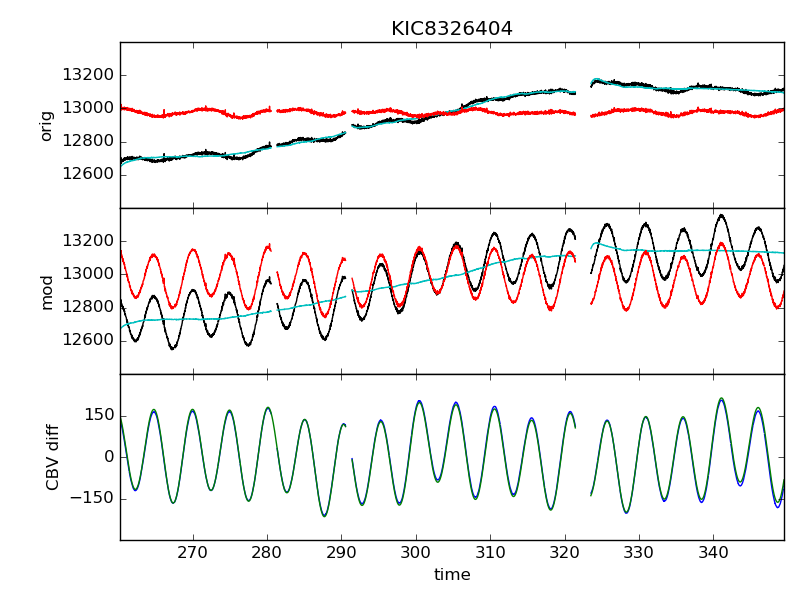} \hfill
  \includegraphics[width=0.49\textwidth]{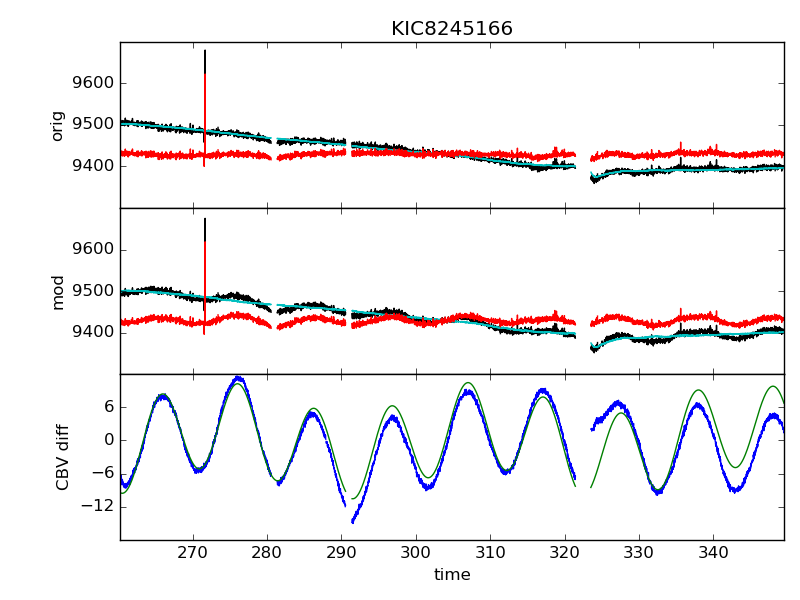}
  \caption{Injection test examples (Q3, modout 17.2). Top: original
    and corrected light curve (black and red, respectively) with the correction
    shown in cyan. Middle: same after injecting the simulated
    signal. Bottom: injected and recovered signals (\modif{green and blue}, respectively.}
  \label{fig:inject_ex}
\end{figure*}

Our prescription for selecting the number of CBVs used, as described in
the previous section, is intended to remove systematics as effectively
as possible while minimising the risk of
over-fitting, i.e. removing astrophysical variability, and of
introducing extra noise into the light curves due to the noisy nature
of the CBVs themselves. To test the extent to which our systematics
removal method affects stellar variability or transit signals, we now perform a series of
injection tests. Specifically, we are interested in variability caused
by rotational modulation of surface inhomogeneities such as
star-spots, since this is a powerful diagnostic of stellar rotation
rates and hence angular momentum evolution, as well as simulated planetary transits.

\begin{figure*}
  \centering
  \includegraphics[width=0.48\textwidth]{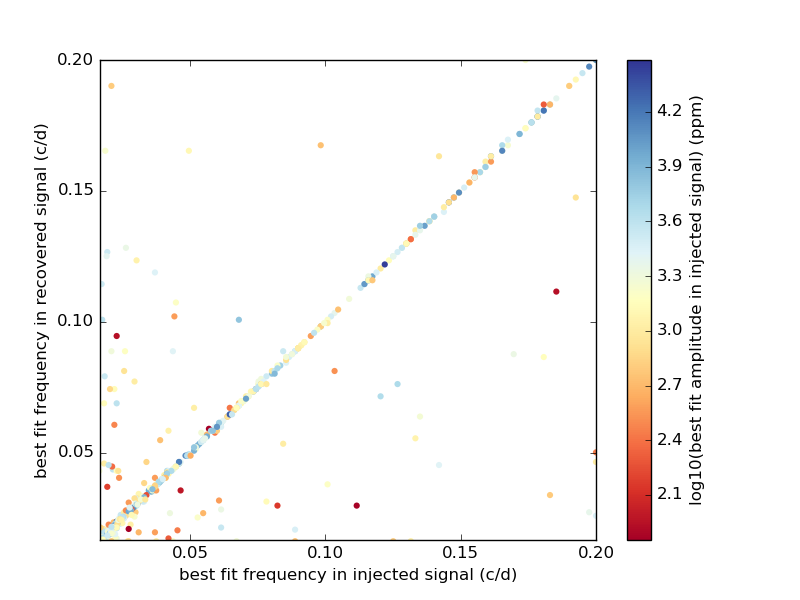} \hfill
  \includegraphics[width=0.48\textwidth]{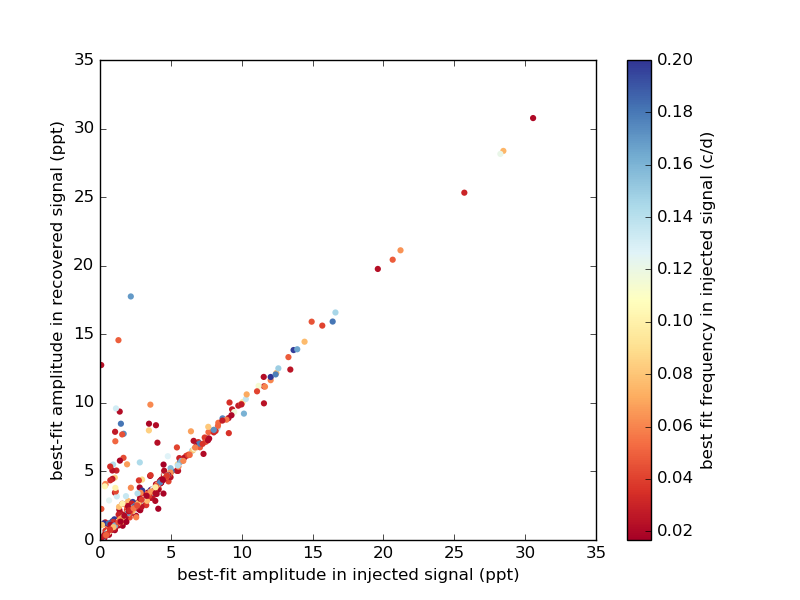} 
  \caption{\modif{Sinusoid injection test results. The left panel shows the frequency of the best-fitting sinusoids in the recovered versus the injected signal, colour coded according to the amplitude of the best-fitting sinusoid in the injected signal. The right panel shows the amplitudes colour-coded according to frequency.}}
  \label{fig:inject_stat}
\end{figure*}

\begin{figure*}
  \centering
  \includegraphics[width=\textwidth]{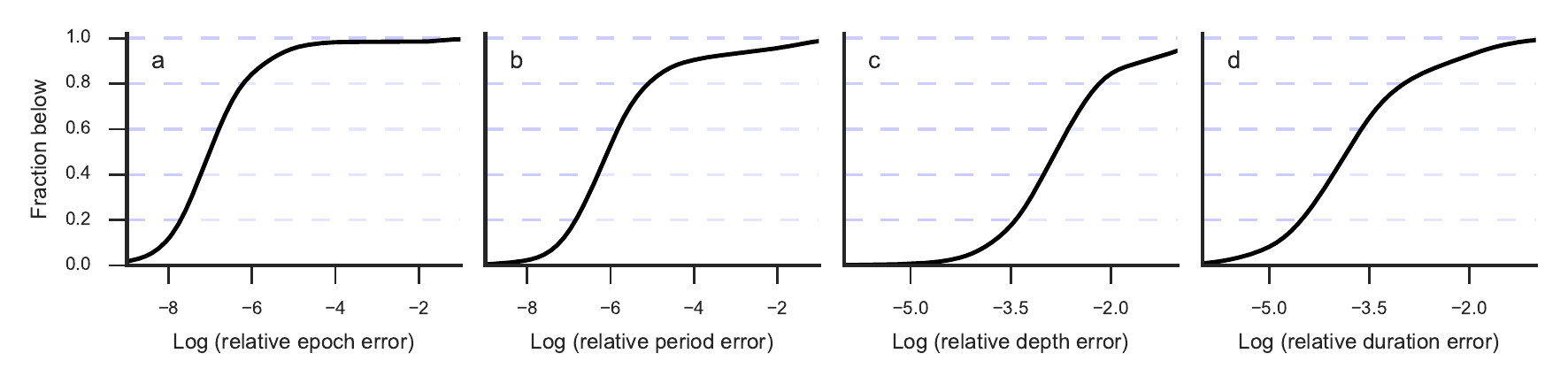} 
  \caption{Transit injection test results as cumulative distributions of the relative errors (the absolute difference between the true and the recovered value divided by the true value) for the recovered transit epoch (a), orbital period (b), transit depth (c), and transit duration (d).}
  \label{fig:inject_stat_transit}
\end{figure*}

\subsubsection*{Rotation-like signals}
We simulate rotation-like signals, consisting of between 1 and 5 co-added
sinusoidal variations, with periods randomly drawn from a log uniform
distribution ranging from 5 to 60 days, and amplitudes drawn from a
log normal distribution with mean $10^{-3}$ and standard deviation 0.5
dex. These were added to 500 randomly selected light curves in each channel and quarter, which were then corrected for systematics as described in the
previous section. The difference between the corrected light curves
with and without injected signal, hereafter referred to as the
\emph{recovered} signal, is then compared to the injected
signal itself: any discrepancies arise because the correction is affecting
the injected signal. Figure~\ref{fig:inject_ex} illustrates this process
for a few example light curves. Of course, the light curves into which the
simulated signals are injected already contain astrophysical
variability, which itself may have been affected by the
correction, and this may contribute to the differences between the
injected and recovered signals. However, we opted for this approach
rather than attempting to generate light curves with simulated stellar
signals \emph{and} systematics, because we do not have a good
generative model for the latter. 

We quantify the effect of the correction on the simulated signals in several ways":
\begin{itemize}
\item \modif{by measuring the Pearson correlation coefficient $P$ between the injected
and recovered signals. We found that $P>0.9$ for 83\% the time;}
\item \modif{by measuring how much high-frequency noise the ARC2 correction adds to the recovered signal. We define the added noise as the quadrature difference between the point-to-point scatter in the recovered versus the injected signal (the latter is of course noise free but still has non-zero \emph{measured} point to point scatter because of its discrete sampling). We found that the added scatter was $<100$\,ppm 92\% of the time;}
\item \modif{by comparing the dominant frequency and amplitude in the injected and recovered signals. We did this by performing a simple least squares fit of a single sinusoid over a grid of equally spaced trial frequencies and extracting the frequency and amplitude of the best-fit sinusoid. The results are shown for a representative channel\footnote{The plots shown are for
modout 17.2 in Q3, but the same tests were performed for all 4 of the
test modouts in Q3--Q6 inclusive, and gave similar results.} and quarter in Figure~\ref{fig:inject_stat}. Overall, the preservation of the signal is excellent: most points tightly cluster around the one-to-one line, and in over 80\% of the cases, the frequency error is $<0.05$ and the amplitude error $<1$\,ppt. We visually examined a subset of the remainder and found that, for most of them, the sine-fitting process had identified a different component of the multi-sine signal in the injected and recovered versions, meaning that the relative amplitudes of the different sinusoids had been significantly altered, but the individual frequencies had not. In most of these cases, the dominant sinusoid in the injected signal had relatively low frequency and/or amplitude, as one would expect.}
\end{itemize}

\subsubsection*{Transit signals}
We carry out the transit-signal injection tests likewise to the rotation-signal injection tests. The orbital periods are drawn randomly from a log uniform distribution ranging from 5 to 60 days, the transit depths are drawn from a log normal distribution with mean $10^{-3}$ and standard deviation 0.5 dex, as with the rotation-signal injection test, and the transit epoch is uniformly distributed. The orbital inclination is fixed to $\frac{\pi}{2}$, semi-major axis is calculated assuming stellar density of 1.5~g$\;$cm$^{-3}$, and the transit duration is derived from the semi-major axis and period assuming a circular orbit. The transit signals are generated using \textsc{PyTransit}\footnote{\url{https://github.com/hpparvi/PyTransit}} \citep{Parviainen2015} with six subsamples per long cadence exposure.

\begin{figure}
  \centering
  \includegraphics[width=\linewidth]{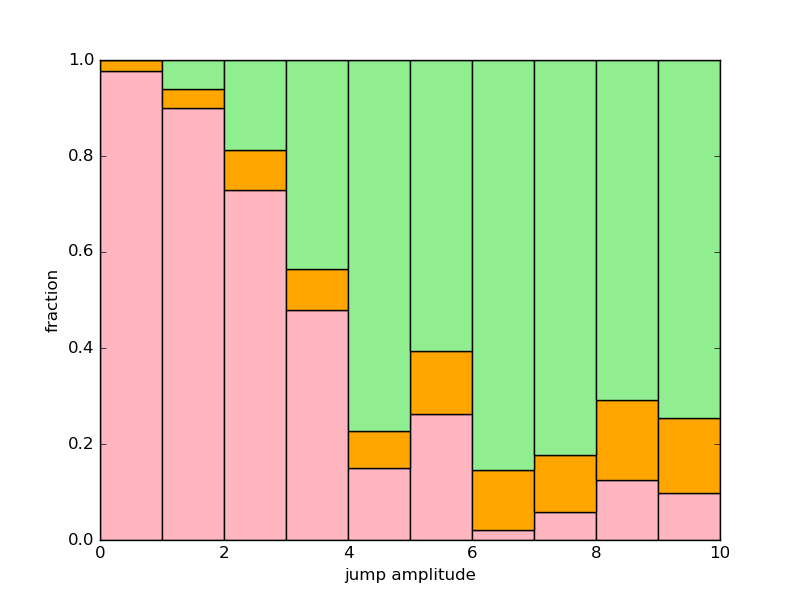} 
  \caption{\modif{Jump injection test results. The $x$-axis shows the amplitude of the injected jumps relative to the high-frequency scatter in the light curves. In each bin, the pink, orange and green bars represent the fraction of injected jumps which were respectively missed (not detected), misclassified (and hence not corrected) and corrected.}}
  \label{fig:inj_jump}
\end{figure}

We add the transit signals to a sample of 500 light curves, carry out the detrending, and fit a transit model to the recovered transit signals (the difference between the detrended light curves with and \modif{without} the injected signal), similarly to the rotation-like signal injection tests. The transit model is fitted using a Nelder-Mead optimiser with the injected signal parameters as starting values.

The results from the transit injection test are shown in Fig.~\ref{fig:inject_stat_transit} as cumulative distributions of the relative errors (absolute difference of the true and recovered value divided by the true value) for the transit epoch, orbital period, transit depth, and transit duration. As with the rotation-signal injection test, we see that the detrending has only a minimal effect on the recovered parameters.

\subsubsection{\modif{Testing the jump correction}}

\modif{We also used the transit injection tests to check whether the jump correction sometimes unintentionally removes planetary transits. To do this, we check what fraction of the time one of the injected transit was detected as a discontinuity, and misclassified as a jump. This happened on average in 10\% of the light curves, but in each light curve the jump correction never removed more than of the injected transit events. Overall, about 2\% of the injected transits were removed.}

\modif{We also performed an additional set of injection tests using simple step function discontinuities, to test how effective the jump correction was at detecting and correcting jumps. The discontinuities were injected at random locations, with (positive or negative) amplitudes ranging from 0 to $10\,\sigma$ (where $\sigma$ is the point-to-point scatter of the light curve). The jump correction was then applied and we checked which of the injected jumps were correctly detected, classified and corrected. The results are shown in Figure~\ref{fig:inj_jump}. The performance improves with jump amplitude, as one would expect, but the overall sensitivity is lower than expected, and it is puzzling that even at the largest amplitude a significant fraction (10\%) of the injected jumps are still missed or misclassified. Most of these cases occur where a simulated jump has been inserted close to a pre-existing, real discontinuity. Our approach relies on the assumption that there are relatively few discontinuities per light curve, so that the likelihood of encountering two of them close together is very low. This assumption is appropriate for real \emph{Kepler} light curves, but is not always satisfied in our injection tests. Therefore, the results of these tests should be taken as a lower limit to the performance of our discontinuity detection and classification method.} 

\subsection{Comparison to PDC-MAP}
\label{sec:comp}

Figures~\ref{fig:ex_pdc} and \ref{fig:sig_ran_comp} were introduced in Section~\ref{sec:pdc} when discussing the PDC pipeline. We now return to them for the purposes of comparing the \modif{PDC} to our new ARC2 pipeline. The most important remark, whether comparing individual examples as in Figure~\ref{fig:ex_pdc}, or statistics over a large ensemble of light curves, as in Figure~\ref{fig:sig_ran_comp}, is how similar the PDC and ARC2 corrections are. This is not altogether surprising,
since both corrections are based on the same set of basis vectors, but
it is reassuring: it indicates that the different choices of priors do
not usually have too strong an effect on the results. It also means that we have succeeded in one of our initial aims, namely to match the overall performance of the PDC pipeline. 

Both corrections do a good job of preserving, or even reducing, the high frequency noise level. However, for
a subset of the stars, the PDC correction a) reduces the range $R$ significantly more, and b) injects significantly more
noise on transit timescales, than the ARC2 one, as illustrated by the bottom two rows of Figure~\ref{fig:ex_pdc}, and by the middle and right panels of the bottom row of Figure~\ref{fig:sig_ran_comp}. In
quantitative terms, for the 4 test channels in Q3, the median value of $R_{\rm PDC}/R_{\rm ARC2}$ was 0.94, and the PDC pipeline increased the 6.5-h CDPP by $>1.1$ in 5.7\% of the stars, compared to only 0.3\% of the stars for the ARC2 pipeline. This suggests that the ARC2 correction is marginally more robust than the PDC one, though the differences are minimal. 

As an aside, we note that comparing $\sigma$ and the 6.5-h CDPP before and after
correction can be used as a useful diagnostic of problematic light
curves. For example, when the ARC2 correction increases $\sigma$ by
more than 10\%, visual examination of the light curves often reveals
abnormal behaviour, apparently associated with image artefacts, stars
located near the edge of a CCD, or very high proper motion stars
(where the fraction of the flux collected within the photometric
aperture might change significantly during a quarter). 

\section{Summary and conclusions}

The injection tests and comparison to the PDC-MAP data detailed in the previous section indicate that the approach we took succeeded with respect to our key objectives: developing an open-source alternative to the PDC-MAP pipeline, which matches (or even slightly improves on) the latter's performance in terms of systematics suppression, while providing demonstrably excellent preservation of astrophysical signals such as rotational modulation of star spots and planetary transits.   

Since we have not demonstrated any significant improvement in performance over the PDC-MAP pipeline, and the PDC-MAP light curves are publicly available for the entire Kepler dataset, it might be reasonable to ask whether publishing an alternative pipeline is worthwhile. The key difference is that our pipeline code is publicly available, in contrast to the Kepler pipeline. This makes it straightforward for users to test the effect of the systematics correction on whichever signals they are most interested in, using, for example, injection tests such as those described in Section~\ref{sec:inject}. This ability is key for any study that aims to draw statistical inferences, for example measuring the distribution of rotation periods and amplitudes for stellar signals, or period and radius ratios for planetary transits.

\modif{Furthermore, the detailed comparison of the two methods, which we have presented in this paper, helps validate both of them. On the whole, we have shown that both are performing well, and if they have limitations, these are mostly common to the two methods. Having two sets of publicly available light curves (or the ability to generate them easily, in the case of the ARC2) is also valuable. If the two versions of a given light curves differ significantly, or if subsequent analysis of the two light curve versions yields different results, this should act as a warning sign to users that something may be amiss. By flagging and examining these cases, the community may eventually help to identify the source of the discrepancy and to improve the performance of both methods.}

\section*{Acknowledgements}
\modif{SA was supported by grant ST/K00106X/1 from the UK Science and Technology Facilities Council, and HP by grant RPG-2012-661 from the Leverhulme Trust. The authors are grateful to the anonymous referee for their careful reading of the manuscript and helpful suggestions.}

\label{lastpage}

\bibliography{arc2}

\begin{thebibliography}{}

\bibitem[\protect\citeauthoryear{{Aigrain}, {Parviainen} \& {Pope}}{{Aigrain}
  et~al.}{2016}]{aig+16}
{Aigrain} S.,  {Parviainen} H.,    {Pope} B.~J.~S.,  2016, MNRAS, 459, 2408

\bibitem[\protect\citeauthoryear{{Bakos}, {Kov{\'a}cs}, {Torres}, {Fischer},
  {Latham}, {Noyes}, {Sasselov}, {Mazeh}, {Shporer}, {Butler}, {Stefanik},
  {Fern{\'a}ndez} \& {et al.}}{{Bakos} et~al.}{2007}]{bak+07}
{Bakos} G.~{\'A}.,  {Kov{\'a}cs} G.,  {Torres} G.,  {Fischer} D.~A.,  {Latham}
  D.~W.,  {Noyes} R.~W.,  {Sasselov} D.~D.,  {Mazeh} T.,  {Shporer} A.,
  {Butler} R.~P.,  {Stefanik} R.~P.,  {Fern{\'a}ndez} J.~M.,    {et al.} 2007,
  ApJ, 670, 826

\bibitem[\protect\citeauthoryear{{Basri}, {Walkowicz}, {Batalha}, {Gilliland},
  {Jenkins}, {Borucki}, {Koch}, {Caldwell}, {Dupree}, {Latham}, {Marcy},
  {Meibom} \& {Brown}}{{Basri} et~al.}{2011}]{bas+10b}
{Basri} G.,  {Walkowicz} L.~M.,  {Batalha} N.,  {Gilliland} R.~L.,  {Jenkins}
  J.,  {Borucki} W.~J.,  {Koch} D.,  {Caldwell} D.,  {Dupree} A.~K.,  {Latham}
  D.~W.,  {Marcy} G.~W.,  {Meibom} S.,    {Brown} T.,  2011, AJ, 141, 20

\bibitem[\protect\citeauthoryear{{Basri}, {Walkowicz}, {Batalha}, {Gilliland},
  {Jenkins}, {Borucki}, {Koch}, {Caldwell}, {Dupree}, {Latham}, {Meibom},
  {Howell} \& {Brown}}{{Basri} et~al.}{2010}]{bas+10a}
{Basri} G.,  {Walkowicz} L.~M.,  {Batalha} N.,  {Gilliland} R.~L.,  {Jenkins}
  J.,  {Borucki} W.~J.,  {Koch} D.,  {Caldwell} D.,  {Dupree} A.~K.,  {Latham}
  D.~W.,  {Meibom} S.,  {Howell} S.,    {Brown} T.,  2010, ApJL, 713, L155

\bibitem[\protect\citeauthoryear{{Fanelli}, {Jenkins}, {Haas} \&
  {Gautier}}{{Fanelli} et~al.}{2011}]{KeplerDataProcHandbook}
{Fanelli} M.,  {Jenkins} J.,  {Haas} M.,    {Gautier} T., , 2011, Kepler Data
  Processing Handbook

\bibitem[\protect\citeauthoryear{{Huang}, {Shen}, {Long}, {Wu}, {Shih},
  {Zheng}, {Yen}, {Tung} \& {Liu}}{{Huang} et~al.}{1998}]{hua98}
{Huang} N.~E.,  {Shen} Z.,  {Long} S.~R.,  {Wu} M.~C.,  {Shih} H.~H.,  {Zheng}
  Q.,  {Yen} N.-C.,  {Tung} C.~C.,    {Liu} H.~H.,  1998, Proceedings of the
  Royal Society of London Series A, 454, 903

\bibitem[\protect\citeauthoryear{{Jenkins}, {Caldwell}, {Chandrasekaran},
  {Twicken}, {Bryson}, {Quintana}, {Clarke}, {Li}, {Allen}, {Tenenbaum}, {Wu},
  {Klaus}, {Middour} \& {et al.}}{{Jenkins} et~al.}{2010}]{jen+10}
{Jenkins} J.~M.,  {Caldwell} D.~A.,  {Chandrasekaran} H.,  {Twicken} J.~D.,
  {Bryson} S.~T.,  {Quintana} E.~V.,  {Clarke} B.~D.,  {Li} J.,  {Allen} C.,
  {Tenenbaum} P.,  {Wu} H.,  {Klaus} T.~C.,  {Middour} C.~K.,    {et al.} 2010,
  ApJL, 713, L87

\bibitem[\protect\citeauthoryear{{Kov{\'a}cs}, {Bakos} \& {Noyes}}{{Kov{\'a}cs}
  et~al.}{2005}]{kov+05}
{Kov{\'a}cs} G.,  {Bakos} G.,    {Noyes} R.~W.,  2005, MNRAS, 356, 557

\bibitem[\protect\citeauthoryear{{McQuillan}, {Aigrain} \& {Mazeh}}{{McQuillan}
  et~al.}{013a}]{mcq+13a}
{McQuillan} A.,  {Aigrain} S.,    {Mazeh} T.,  2013a, MNRAS, 432, 1203

\bibitem[\protect\citeauthoryear{{McQuillan}, {Mazeh} \& {Aigrain}}{{McQuillan}
  et~al.}{013b}]{mcq+13b}
{McQuillan} A.,  {Mazeh} T.,    {Aigrain} S.,  2013b, ApJL, 775, L11

\bibitem[\protect\citeauthoryear{{McQuillan}, {Mazeh} \& {Aigrain}}{{McQuillan}
  et~al.}{2014}]{mcq+14}
{McQuillan} A.,  {Mazeh} T.,    {Aigrain} S.,  2014, ApJ, in press

\bibitem[\protect\citeauthoryear{{Nielsen}, {Gizon}, {Schunker} \&
  {Karoff}}{{Nielsen} et~al.}{2013}]{nie+13}
{Nielsen} M.~B.,  {Gizon} L.,  {Schunker} H.,    {Karoff} C.,  2013, A\&A, 557,
  L10

\bibitem[\protect\citeauthoryear{Parviainen}{Parviainen}{2015}]{Parviainen2015}
Parviainen H.,  2015, MNRAS, 450, 3233

\bibitem[\protect\citeauthoryear{{Reinhold}, {Reiners} \& {Basri}}{{Reinhold}
  et~al.}{2013}]{rei+13}
{Reinhold} T.,  {Reiners} A.,    {Basri} G.,  2013, A\&A, 560, A4

\bibitem[\protect\citeauthoryear{{Roberts}, {McQuillan}, {Reece} \&
  {Aigrain}}{{Roberts} et~al.}{2013}]{rob+13}
{Roberts} S.,  {McQuillan} A.,  {Reece} S.,    {Aigrain} S.,  2013, MNRAS, 435,
  3639

\bibitem[\protect\citeauthoryear{{Smith}, {Stumpe}, {Van Cleve}, {Jenkins},
  {Barclay}, {Fanelli}, {Girouard}, {Kolodziejczak}, {McCauliff}, {Morris} \&
  {Twicken}}{{Smith} et~al.}{2012}]{smi+12}
{Smith} J.~C.,  {Stumpe} M.~C.,  {Van Cleve} J.~E.,  {Jenkins} J.~M.,
  {Barclay} T.~S.,  {Fanelli} M.~N.,  {Girouard} F.~R.,  {Kolodziejczak} J.~J.,
   {McCauliff} S.~D.,  {Morris} R.~L.,    {Twicken} J.~D.,  2012, PASP, 124,
  1000

\bibitem[\protect\citeauthoryear{{Stumpe}, {Smith}, {Catanzarite}, {Van Cleve},
  {Jenkins}, {Twicken} \& {Girouard}}{{Stumpe} et~al.}{2014}]{stu+14}
{Stumpe} M.~C.,  {Smith} J.~C.,  {Catanzarite} J.~H.,  {Van Cleve} J.~E.,
  {Jenkins} J.~M.,  {Twicken} J.~D.,    {Girouard} F.~R.,  2014, PASP, 126, 100

\bibitem[\protect\citeauthoryear{{Stumpe}, {Smith}, {Van Cleve}, {Twicken},
  {Barclay}, {Fanelli}, {Girouard}, {Jenkins}, {Kolodziejczak}, {McCauliff} \&
  {Morris}}{{Stumpe} et~al.}{012a}]{stu+12a}
{Stumpe} M.~C.,  {Smith} J.~C.,  {Van Cleve} J.~E.,  {Twicken} J.~D.,
  {Barclay} T.~S.,  {Fanelli} M.~N.,  {Girouard} F.~R.,  {Jenkins} J.~M.,
  {Kolodziejczak} J.~J.,  {McCauliff} S.~D.,    {Morris} R.~L.,  2012a, PASP,
  124, 985

\bibitem[\protect\citeauthoryear{{Tamuz}, {Mazeh} \& {Zucker}}{{Tamuz}
  et~al.}{2005}]{tam+05}
{Tamuz} O.,  {Mazeh} T.,    {Zucker} S.,  2005, MNRAS, 356, 1466

\bibitem[\protect\citeauthoryear{{Thompson}, {Fraquelli}, {Van Cleve} \&
  {Caldwell}}{{Thompson} et~al.}{2016}]{KeplerArchiveManual}
{Thompson} S.,  {Fraquelli} D.,  {Van Cleve} J.~E.,    {Caldwell} D.~A., ,
  2016, Kepler Archive Manual

\bibitem[\protect\citeauthoryear{{Twicken}, {Chandrasekaran}, {Jenkins},
  {Gunter}, {Girouard} \& {Klaus}}{{Twicken} et~al.}{010a}]{twi+10a}
{Twicken} J.~D.,  {Chandrasekaran} H.,  {Jenkins} J.~M.,  {Gunter} J.~P.,
  {Girouard} F.,    {Klaus} T.~C.,  2010a, in Society of Photo-Optical
  Instrumentation Engineers (SPIE) Conference Series Vol.~7740 of Society of
  Photo-Optical Instrumentation Engineers (SPIE) Conference Series, {Presearch
  data conditioning in the Kepler Science Operations Center pipeline}

\bibitem[\protect\citeauthoryear{{Twicken}, {Clarke}, {Bryson}, {Tenenbaum},
  {Wu}, {Jenkins}, {Girouard} \& {Klaus}}{{Twicken} et~al.}{010b}]{twi+10b}
{Twicken} J.~D.,  {Clarke} B.~D.,  {Bryson} S.~T.,  {Tenenbaum} P.,  {Wu} H.,
  {Jenkins} J.~M.,  {Girouard} F.,    {Klaus} T.~C.,  2010b, in Society of
  Photo-Optical Instrumentation Engineers (SPIE) Conference Series Vol.~7740 of
  Society of Photo-Optical Instrumentation Engineers (SPIE) Conference Series,
  {Photometric analysis in the Kepler Science Operations Center pipeline}

\bibitem[\protect\citeauthoryear{{van Cleve}, {Caldwell}, {Thompson}, {Haas},
  {Koch} \& {Borucki}}{{van Cleve} et~al.}{2009}]{KeplerInstrumentHandbook}
{van Cleve} J.,  {Caldwell} D.,  {Thompson} R.,  {Haas} M.,  {Koch} D.,
  {Borucki} W., , 2009, Kepler Instrument Handbook

\end{thebibliography}
\bibliographystyle{mn2e}

\end{document}